\documentclass[12pt]{article}%
\usepackage{ amsmath, graphics, rotating}%

\begin{document}
\begin{center}{\large\bf Existence of Antiparticles as an Indication of Finiteness of Nature} \end{center}
\vskip 1em \begin{center} {\large Felix M. Lev} \end{center}
\vskip 1em \begin{center} {\it Artwork Conversion Software Inc.,
1201 Morningside Drive, Manhattan Beach, CA 90266, USA
(Email:  felixlev314@gmail.com)} \end{center}
\vskip 1em

\begin{flushleft}{\it Abstract:}\end{flushleft} 
It is shown that in a quantum theory over a Galois field, the famous Dirac's result about antiparticles 
is generalized such that a particle and its antiparticle are already combined at the level of 
irreducible representations of the symmetry algebra without assuming the existence of a local covariant 
equation. We argue that the very existence of antiparticles
is a strong indication that nature is described by a finite field rather than by complex numbers. 

\begin{flushleft} PACS: 02.10.De, 03.65.Ta; MSC: 20C35, 81R05\end{flushleft}

\begin{flushleft} Keywords: particle-antiparticle, Galois fields, Lie algebras, modular representations\end{flushleft}

\section{Problem statement}
\label{S1}

A well-known fact of particle physics is that a particle and its antiparticle have equal masses.
The explanation of this fact in quantum field theory (QFT) follows.

Irreducible representations (IRs) of the Poincare and anti-de Sitter (AdS) algebras by Hermitian operators
used for describing elementary particles have the property that for each IR the Hamiltonian is either positive definite or negative definite. In the first case, the energy has the spectrum in the range $[mass,\infty)$, while in the second case it has the spectrum in the range $(-\infty,-mass]$. 

However, for constructing Lagrangians one needs to work not with IRs but with local fields
satisfying covariant equations (e.g. the Klein-Gordon equation, the Dirac equation etc.). Those fields
are described by non-unitary representations of the Poincare or AdS groups induced from the Lorentz group. 
Each local field 
is a combination of two IRs with positive and negative energies called a particle and its antiparticle,
respectively. Then, as follows from the CPT theorem, a particle and its antiparticle
have the same masses. The problem of negative energies is then solved 
by quantization, after which the energies of both, the particle and its antiparticle become positive definite. 

One might pose the following question. If locality is only approximate then the masses of
a particle and its antiparticle should remain equal or can differ each other? This question
is legitimate because the physical meaning of locality is not quite clear.
The matter is that since local fields are described by non-unitary representations, their
probabilistic interpretation is problematic. As shown by Pauli \cite{Pauli2},
in the case of fields with an integer spin there is no subspace where the spectrum of the charge operator has a definite sign
while in the case of fields with a half-integer spin there is no subspace where the spectrum of the energy operator has a definite sign.
Local fields is only an auxiliary tool for
constructing operators describing unitary representations of a system as a whole (momentum, energy, 
angular momentum etc.).

In the present paper we investigate the status of particles and antiparticles in a quantum theory
over a Galois field (GFQT) proposed first in Refs. \cite{lev2,lev3}. The motivation and a detailed
description of GFQT can be also found in Refs. \cite{PRD,monograph}. In GFQT quantum states are
elements of linear spaces over a Galois field of characteristic $p$ and operators of physical
quantities are linear operators in such spaces. Since any Galois field is finite, in GFQT infinities
cannot exist in principle. At the same time, when $p$ is rather large, GFQT recovers predictions
of standard quantum theory. 

The idea of correspondence between GFQT and standard theory follows. If $p$ is prime then the Galois field $F_p$ with $p$ elements can be
represented as a set of elements $\{0,\pm i\}$ $(i=1,2,...(p-1)/2)$. Let $f$ be a function from $F_p$ to the ring of integers $Z$ such that
$f(a)$ in $Z$ has the same notation in $Z$ as $a$ in $F_p$. Then for elements $a\in F_p$ such that $|f(a)|\ll p$, 
addition, subtraction and multiplication are the same as in $Z$. In other words, for
such elements we do not feel the existence of $p$. Indeed, if the elements $a_j$ $(j=1,2,...n)$ are such that
$|f(a_j)|<[(p-1)/2]^{1/n}$ then 
$$f(\sum_{j=1}^n a_j)=\sum_{j=1}^n f(a_j),\,\, f(\prod_{j=1}^n a_j)=\prod_{j=1}^n f(a_j)$$
which shows that if $F_p$ is treated as a ring then $f$ is a local isomorphism between $F_p$ and $Z$. When $p$ increases, 
the bigger and bigger part of $F_p$
becomes the same as $Z$. This important observation implies that {\it $Z$ can be treated as a special case of 
$F_p$ in the formal limit $p\to\infty$}.

In the general case, division in $F_p$ is not the same as in standard mathematics. For example, 1/2 in $F_p$ equals
$(p+1)/2$, i.e. a very large number if $p$ is large. However, this does not mean that mathematics modulo $p$
cannot describe physics. The matter is that spaces in quantum theory are projective. 

By analogy with standard quantum theory, it is natural to define the elementary particle in GFQT as
a system described by an IR of a Lie algebra over a Galois field. Representations of Lie algebras in spaces with nonzero characteristic are called 
modular representations. There exists a well developed theory of such representations. One of the known results is the Zassenhaus theorem \cite{Zass} that any modular IR is finite dimensional. 

It is well-known that any Galois field contains $p^k$ elements where $p$ is prime and $k$ is natural.
We use $F_{p^k}$ to denote such fields. In standard theory one considers representations
of real Lie algebras in complex Hilbert spaces. Modular analogs of such representations are
representations of Lie algebras over $F_p$ in spaces over $F_{p^2}$. However, the following
remark is in order.

Consider when the elements of $F_{p^2}$ can be represented as $a+bi$ where $a,b\in F_p$ and $i$ is a formal element such that $i^2=-1$.
The division in $F_{p^2}$ can be defined as $(a+bi)^{-1}=(a-bi)/(a^2+b^2)$ if $(a^2+b^2)=0$ in $F_p$ implies that $a=0$ and $b=0$.
As explained in textbooks on number theory, this is the case only if $p=3\,\, (mod\,\, 4)$. Therefore $F_{p^2}$ with such values of $p$ can be treated as analogs of complex numbers.

Since we treat standard theory as a special case of GFQT in the formal limit $p\to\infty$, it is desirable not to postulate
that GFQT is based on $F_{p^2}$ with $p=3\,\,(mod \,4)$ because standard theory is based on complex numbers
but vice versa, explain the fact that standard 
theory is based on complex numbers since GFQT is based 
on $F_{p^2}$. Therefore we should find a motivation for
the choice of $F_{p^2}$ with $p=3\,\,(mod \,4)$. Arguments in favor of such a choice
are discussed in Refs. \cite{lev3,complex,monograph}. In this paper we will not use the
restriction that the representation space is over $F_{p^2}$ with $p=3\,\, (mod\,\, 4)$ and will consider a general
case when it is over $F_{p^k}$.

In standard quantum theory, Poincare symmetry is a special case of de Sitter (dS) or AdS
symmetries in the procedure called contraction. As shown in Refs. \cite{lev3,monograph},
in GFQT there is no analog of Poincare symmetry but analogs of dS and AdS symmetries are well defined.
In the present paper we consider modular analogs of IRs of the AdS algebra while the case of IRs
of the dS algebra is mentioned in Sec. \ref{discussion}. 

The paper is organized as follows. In Sec. \ref{S2} we explicitly construct modular IRs
of the sp(2) algebras. Such IRs play an important auxiliary role for constructing 
modular IRs of the AdS algebra in Sec. \ref{S3}. The results show that the status of particles
and antiparticles in GFQT considerably differs from that in standard theory. Finally
Sec. \ref{discussion} is a discussion.   

\section{Modular IRs of the sp(2) algebra}
\label{S2}

The key role in constructing modular IRs of the so(2,3) algebra is played by modular IRs of the sp(2) 
subalgebra. They are described by a set of operators $(a',a",h)$ satisfying the commutation relations
\begin{equation}
[h,a']=-2a',\quad [h,a"]=2a",\quad [a',a"]=h
\label{2}
\end{equation}
The  Casimir operator of the second order for the algebra (\ref{2}) has the form
\begin{equation}
K=h^2-2h-4a"a'=h^2+2h-4a'a"
\label{3}
\end{equation}

We first consider representations with the vector $e_0$ such that
\begin{equation}
a'e_0=0,\quad he_0=q_0e_0
\label{4}
\end{equation}
where $q_0\in F_p$ and $f(q_0) > 0$.
Denote $e_n =(a")^ne_0$.
Then it follows from Eqs. (\ref{3}) and (\ref{4}),
that \begin{equation}
he_n=(q_0+2n)e_n,\quad Ke_n=q_0(q_0-2)e_n,
\label{5}
\end{equation}
\begin{equation}
a'a"e_n=(n+1)(q_0+n)e_n
\label{6}
\end{equation}

One can consider analogous representations in standard theory. Then $q_0$ is a positive real
number, $n=0,1,2,...\infty$ and the elements $e_n$ form 
a basis of the IR. In this case $e_0$ is a vector with a minimum eigenvalue of the
operator $h$ (minimum weight) and there are no vectors with the maximum weight.
The operator $h$ is positive definite
and bounded below by the quantity $q_0$. For these
reasons the above modular IRs can be treated as
modular analogs of such standard IRs that $h$ is positive definite.

Analogously, one can construct modular IRs starting from
the element $e_0'$ such that
\begin{equation}
a"e_0'=0,\quad he_0'=-q_0e_0'
\label{7}
\end{equation}
and the elements $e_n'$ can be defined as $e_n'=(a')^ne_0'$.
Such modular IRs are analogs of standard IRs where $h$ is negative definite. 
However, in the modular case one can easily prove the following statement.

{\bf Theorem 1}: {\it Eqs. (\ref{4}) and (\ref{7}) define the same IR with the dimension $p-q_0+1$}.

{\it Proof}. The set $(e_0,e_1,...e_N)$ will be a basis of
IR if $a"e_i\neq 0$
for $i<N$ and $a"e_N=0$. These conditions
must be compatible with $a'a"e_N=0$.
As follows from
Eq. (\ref{6}), $N$ is defined by the condition 
$q_0+N=0$ in $F_p$. As a result, if
$q_0$ is one of the numbers $1,...p-1$
then $N=p-q_0$ and the dimension of IR equals
$p-q_0+1$ (in agreement with the Zassenhaus
theorem \cite{Zass}). The element $e_N$ satisfies Eq. (\ref{7}) and therefore
it can be identified with $e_0'$.

\section{Modular IRs of the so(2,3) algebra}
\label{S3}

Standard IRs of the AdS so(2,3) algebra relevant for 
describing elementary particles have been 
considered by many authors. The description
in this section is a combination of two elegant ones
given in Ref. \cite{Evans} for standard IRs and Ref.
\cite{Braden} for modular IRs. 
In standard theory the representation
operators of the so(2,3) algebra in units
$\hbar/2=c=1$ are given by 
\begin{equation}
[M^{ab},M^{cd}]=-2i (g^{ac}M^{bd}+g^{bd}M^{cd}-
g^{ad}M^{bc}-g^{bc}M^{ad})
\label{8}
\end{equation}
where $a,b,c,d$ take the values 0,1,2,3,4 and $M^{ab}=-M^{ba}$. The diagonal metric tensor
has the components $g^{00}=g^{44}=-g^{11}=-g^{22}=-g^{33}=1$. 
In these units the spin of fermions is odd, and
the spin of bosons is even. If $s$ is the particle spin
then the corresponding IR of the su(2) algebra has
the dimension $s+1$. 

If a modular IR is considered in a linear space over
$F_{p^2}$ with $p=3\,\, (mod\,\, 4)$ then Eq. (\ref{8}) is also valid but in the general case it is
convenient to work with another set of ten operators.
Let $(a_j',a_j",h_j)$ $(j=1,2)$ be two independent sets
of operators satisfying the
commutation relations for the sp(2) algebra
\begin{equation}
[h_j,a_j']=-2a_j',\quad [h_j,a_j"]=2a_j",\quad [a_j',a_j"]=h_j
\label{9}
\end{equation}
The sets are independent in the sense that
for different $j$ they mutually commute with each other.
We denote additional four operators as $b', b",L_+,L_-$.
The operators
$L_3=h_1-h_2,L_+,L_-$ satisfy the commutation relations
of the su(2) algebra
\begin{equation}
[L_3,L_+]=2L_+,\quad [L_3,L_-]=-2L_-,\quad [L_+,L_-]=L_3
\label{10}
\end{equation}
while the other commutation relations are
\begin{eqnarray}
&[a_1',b']=[a_2',b']=[a_1",b"]=[a_2",b"]=[a_1',L_-]=[a_1",L_+]=\nonumber\\
&[a_2',L_+]=[a_2",L_-]=0,\quad [h_j,b']=-b',\quad [h_j,b"]=b",\quad
[h_1,L_{\pm}]=\pm L_{\pm},\nonumber\\
&[h_2,L_{\pm}]=\mp L_{\pm},\quad [b',b"]=h_1+h_2,\quad [b',L_-]=2a_1',\quad [b',L_+]=2a_2',\nonumber\\
&[b",L_-]=-2a_2",\quad [b",L_+]=-2a_1",\quad [a_1',b"]=[b',a_2"]=L_-,\nonumber\\
&[a_2',b"]=[b',a_1"]=L_+,\quad [a_1',L_+]=[a_2',L_-]=b',\nonumber\\
&[a_2",L_+]=[a_1",L_-]=-b"
\label{11}
\end{eqnarray}
At first glance these relations might seem rather
chaotic but in fact they are very natural in the Weyl basis
of the so(2,3) algebra.

In spaces over $F_{p^2}$ with $p=3\,\, (mod\,\, 4)$ the 
relation between the above sets of ten operators is
\begin{eqnarray}
&M_{10}=i(a_1"-a_1'-a_2"+a_2'),\quad M_{14}=a_2"+a_2'-a_1"-a_1',\nonumber\\
&M_{20}=a_1"+a_2"+a_1'+a_2',\quad M_{24}=i(a_1"+a_2"-a_1'-a_2'),\nonumber\\
&M_{12}=L_3,\quad M_{23}=L_++L_-,\quad M_{31}=-i(L_+-L_-),\nonumber\\
&M_{04}=h_1+h_2,\quad M_{34}=b'+b",\quad M_{30}=-i(b"-b')
\label{12}
\end{eqnarray}
which is why the sets are equivalent. The relations
(\ref{9}-\ref{11}) are more general since they can be used
when the representation space is a space over $F_{p^k}$ where 
 $k$ is arbitrary.

We use the basis in which the operators
$(h_j,K_j)$ $(j=1,2)$ are diagonal. Here $K_j$ is the
Casimir operator (\ref{3}) for algebra $(a_j',a_j",h_j)$.
For constructing IRs we need operators relating different
representations of the sp(2)$\times$sp(2) algebra.
By analogy with Refs. \cite{Evans,Braden}, one of the
possible choices is 
\begin{eqnarray}
&A^{++}=b"(h_1-1)(h_2-1)-a_1"L_-(h_2-1)-a_2"L_+(h_1-1)+a_1"a_2"b',\nonumber\\
&A^{+-}=L_+(h_1-1)-a_1"b',\quad A^{-+}=L_-(h_2-1)-a_2"b',\quad A^{--}=b'
\label{13}
\end{eqnarray}
We consider the action of these operators only on the
space of "minimal"
sp(2)$\times$sp(2) vectors, i.e. such vectors $x$ that
$a_j'x=0$ for $j=1,2$, and $x$ is the eigenvector of the
operators $h_j$. If $x$ is a minimal vector such that
$h_jx=\alpha_jx$ then $A^{++}x$ is the minimal
eigenvector of the
operators $h_j$ with the eigenvalues $\alpha_j+1$, $A^{+-}x$ -
with the eigenvalues $(\alpha_1+1,\alpha_2-1)$,
$A^{-+}x$ - with the eigenvalues $(\alpha_1-1,\alpha_2+1)$,
and $A^{--}x$ - with the eigenvalues $\alpha_j-1$.

By analogy with Refs. \cite{Evans,Braden}, we require
the existence of the vector $e_0$ satisfying the conditions
\begin{eqnarray}
&a_j'e_0=b'e_0=L_+e_0=0,\quad h_je_0=q_je_0\quad (j=1,2)
\label{15}
\end{eqnarray}
where $q_j\in F_p$, $f(q_j)>0$ and $f(q_1-q_2)\geq 0$. 
It is well known (see e.g. Refs. \cite{Evans,monograph}) that 
$M^{04}=h_1+h_2$ is the AdS analog of the energy operator.
As follows from
Eqs. (\ref{9}) and (\ref{11}), the operators
$(a_1',a_2',b')$ reduce the AdS energy by two units.
Thus $e_0$ is an analog of the state with the minimum energy
which can be called the rest state, and the spin in our
units is equal to the eigenvalue
of the operator $L_3=h_1-h_2$ in that state. For these
reasons we use $s$ to denote $q_1-q_2$
and $m$ to denote $q_1+q_2$. 
In the standard classification
\cite{Evans}, the massive case is characterized by
the condition $q_2>1$ and the massless case --- by
the condition $q_2=1$. There also exist two
exceptional IRs discovered by Dirac \cite{DiracS}
(Dirac singletons). As shown in Refs. \cite{lev2,monograph},
the modular analog of Dirac singletons is simple
and the massless case has been discussed in
detail in Ref. \cite{tmf}. For these reasons in
the present paper we consider only the massive case.

As follows from the above remarks, the elements
\begin{equation}
e_{nk}=(A^{++})^n(A^{-+})^ke_0
\label{16}
\end{equation}
represent the minimal sp(2)$\times$sp(2) vectors with the
eigenvalues of the operators $h_1$ and $h_2$ equal to
$Q_1(n,k)=q_1+n-k$ and $Q_2(n,k)=q_2+n+k$, respectively.
It can be shown by a direct calculation that
\begin{equation}
A^{--}A^{++}e_{nk}=(n+1)(m+n-2)(q_1+n)(q_2+n-1)e_{nk}
\label{17}
\end{equation}
\begin{equation}
A^{+-}A^{-+}e_{nk}=(k+1)(s-k)(q_1-k-2)(q_2+k-1)e_{nk}
\label{18}
\end{equation}

As follows from these expressions, in the massive
case $k$ can assume only 
the values $0,1,...s$ and in standard theory $n=0,1,...\infty$. 
However, in the modular case the following results are valid.

{\bf Theorem 2:} {\it The full basis of the representation space can 
be chosen in the form
\begin{equation}
e(n_1n_2nk)=(a_1")^{n_1}(a_2")^{n_2}e_{nk}
\label{19}
\end{equation}
The value of $n$ is in the
range $n=0,1,...n_{max}$ where $n_{max}$ is the first number
for which the r.h.s. of Eq. (\ref{17}) becomes zero in $F_p$, i.e. $n_{max}=p+2-m$. 
As follows from {\bf Theorem 1}, Eq. (\ref{9}) and
the properties of the $A$ operators,
\begin{eqnarray}
&n_1=0,1,...N_1(n,k),\quad n_2=0,1,...N_2(n,k),\nonumber\\
&N_1(n,k)=p-q_1-n+k,\quad N_2(n,k)=p-q_2-n-k
\label{20}
\end{eqnarray}
}

As a consequence, the representation is finite 
dimensional in agreement with the Zassenhaus 
theorem \cite{Zass}. Moreover, it is
finite since any Galois field is finite.

In standard Poincare and AdS theories there also exist IRs with
negative energies. They can be constructed by analogy with 
positive energy IRs.
Instead of Eq. (\ref{15}) one can require the existence of the
vector $e_0'$ such that
\begin{eqnarray}
&a_j"e_0'=b"e_0'=L_-e_0'=0,\quad h_je_0'=-q_je_0'\quad (j=1,2)
\label{26}
\end{eqnarray}
where the quantities $q_1,q_2$ are the same as for positive
energy IRs. It is obvious that positive and negative energy
IRs are fully independent since the spectrum of the operator
$M^{04}$ for such IRs is positive and negative, respectively.
However, the following theorem indicates to a crucial difference
between standard theory and GFQT.

{\bf Theorem 3:} {\it The modular analog 
of the positive energy IR characterized by $q_1,q_2$ in
Eq. (\ref{15}), and the modular
analog of the negative energy IR characterized by the same
values of $q_1,q_2$ in Eq. (\ref{26}) represent the same
modular IR.}

{\it Proof}. Let $e_0$ be a vector satisfying Eq. (\ref{15}). Denote
$N_1=p-q_1$ and $N_2=p-q_2$. Our goal is to prove
that the vector $x=(a_1")^{N_1}(a_2")^{N_2}e_0$ satisfies
the conditions
(\ref{26}), i.e. $x$ can be identified with $e_0'$.

As follows from the definition of $N_1,N_2$,
the vector $x$ is the eigenvector of the operators $h_1$
and $h_2$ with the eigenvalues $-q_1$ and $-q_2$,
respectively, and in
addition it satisfies the conditions $a_1"x=a_2"x=0$.
Let us prove that $b"x=0$. Since $b"$ commutes with the
$a_j"$, we can write $b"x$ in the form
\begin{equation}
b"x = (a_1")^{N_1}(a_2")^{N_2}b"e_0
\label{27}
\end{equation}
As follows from Eqs. (\ref{11}) and (\ref{15}),
$a_2'b"e_0=L_+e_0=0$ and $b"e_0$ is the eigenvector
of the operator $h_2$ with the eigenvalue $q_2+1$.
Thus, $b"e_0$ is the minimal vector of the sp(2)
IR which has the dimension $p-q_2=N_2$.
Therefore $(a_2")^{N_2}b"e_0=0$ and $b"x=0$.

The next step is to show that $L_-x=0$.
As follows from Eq. (\ref{11}) and the definition of
$x$,
\begin{equation}
L_-x = (a_1")^{N_1}(a_2")^{N_2}L_-e_0-
N_1(a_1")^{N_1-1}(a_2")^{N_2}b"e_0
\label{28}
\end{equation}
We have already shown that $(a_2")^{N_2}b"e_0=0$,
and hence it suffices to prove that the first term
in the r.h.s. of Eq. (\ref{28}) equals zero. As follows
from Eqs. (\ref{11}) and (\ref{15}), $a_2'L_-e_0=b'e_0=0$
and $L_-e_0$ is the eigenvector of the operator $h_2$ with the
eigenvalue $q_2+1$. Thus, $(a_2")^{N_2}L_-e_0=0$ 
and the proof is completed.

\section{Discussion}
\label{discussion}

The construction in Sec. \ref{S3} applies to both, standard IRs of the so(2,3) algebra and their modular 
analogs. Consider first standard IRs. Here the element $e_0$ defined by Eq. (\ref{15}) is
the state with the minimum energy, i.e. we start from the rest state where, by definition,
energy=mass and the value of the energy is positive. When the representation operators act on $e_0$ one obtains
states with higher and higher energies and the energy spectrum is in the range $[mass,\infty ]$. Analogously the
element $e_0'$ defined by Eq. (\ref{26}) is such that the energy in this state is such that energy=-mass
while the energy spectrum is in the range $[-\infty,-mass]$. As noted in Sec. \ref{S1}, in standard
theory positive and negative energy IRs are called particles and anti-particles, respectively. Here a particle and its antiparticle are different objects because they are described by fully independent IRs.
Then, as noted in Sec. \ref{S1}, a problem arises why a particle and its antiparticle have equal masses.

Let us now discuss what happens in GFQT. We again start from the state $e_0$ and one might think that the
corresponding IR is the modular analog of the standard positive energy IR with the minimum weight. Indeed,
when the operators $A^{++}$ act on $e_0$ we successively obtain states where the energy increases by two units.
However, since the values of the energy now belong not to $Z$ but to $F_p$ then sooner or 
later we will arrive to states where the energy is "negative" (i.e. in the range $[-(p-1)/2,-1]$) 
and finally we will arrive to 
the state where energy=-mass. In mathematical terminology this means that a modular analog of IR with the minimum weight
is simultaneously a modular analog of IR with the maximum weight, while from the point of view of physics, one
modular IR describes a particle and its antiparticle simultaneously. 

As noted in Sec. \ref{S1}, in QFT a question arises that if locality is only approximate then it is not clear whether the notion of antiparticles is exact or approximate and whether they have equal masses. 
At the same time, the above construction shows 
that {\it in GFQT the existence of antiparticles follows from the fact that any Galois field is finite.} 

Consider a simple well-known model of particle theory when electromagnetic and weak interactions are
absent. Then the fact that the proton and the neutron have the same
masses and spins is irrelevant of locality or nonlocality; it is only a consequence of the fact that the proton and 
the neutron belong to the same isotopic multiplet. In other words, they are simply different
states of the same object - the nucleon. We see that in GFQT the
situation is analogous. The fact that a particle and its antiparticle have the
same masses and spins is irrelevant of locality or nonlocality and is simply a consequence of the fact that they are
different states of the same object since they belong to the same IR. 

Note also, that in standard theory, IRs of the dS algebra contain states with both, positive and negative energies
and, as shown in Ref. \cite{JPA}, the only possible interpretation of such IRs is that they describe
a particle and its antiparticle simultaneously.

In summary, while in standard theory the existence of antiparticles depends on additional assumptions, in GFQT it is inevitable. Therefore, {\it the very existence of antiparticles is a strong indication that nature 
is described by a finite field rather than by complex numbers.}

Strictly speaking, the above construction shows that the very notion of particles and antiparticles
is approximate. A set of states where the energy $E$ is such that $f(E)>0$ and $f(E)\ll p$ can be called a particle while a set of states where $f(E)<0$ and $|f(E)|\ll p$ can be called an antiparticle. This situation has far reaching consequences.
A problem also arises how to treat neutral particles where a particle and its antiparticle are the same. Those problems are discussed in Refs. \cite{tmf,JPA,PRD,monograph}.

\end{document}